\documentstyle[aps, prl, twocolumn, floats, epsfig]{revtex}
\begin{document} \draft 
\title{ Overscreening in Hubbard electron systems}
\author{H.-B. Sch\"{u}ttler$^1$, C. Gr\"ober$^2$, H. G. Evertz$^{2,3}$, 
        and W. Hanke$^2$\vskip2mm} 
\address{$^1$ Center for Simulational Physics, Department of Physics
 and Astronomy, University of Georgia, Athens, Georgia 30602 }
\address{$^2$ Institut f\"ur Theoretische Physik, Am Hubland,
Universit\"at W\"urzburg, D-97074 W\"urzburg, Germany }
\address{$^3$Institute for Solid State Physics, University of Tokyo, Tokyo 106-8666, Japan}
\date{May 12, 1998} 
\twocolumn[\hsize\textwidth\columnwidth\hsize\csname @twocolumnfalse\endcsname
\maketitle

\begin{abstract}
We show that doping-induced charge fluctuations 
in strongly correlated Hubbard electron systems near the 
$1\over2$-filled, insulating limit
cause overscreening of the electron-electron Coulomb repulsion.
The resulting attractive screened interaction potential
supports $d_{x^2-y^2}$-pairing with a strongly peaked, doping dependent
pairing strength at lower doping, followed by $s$-wave
pairing at larger doping levels.
\end{abstract} 
]

In the cuprate high-$T_c$ superconductors, 
the strong, on-site Hubbard-$U$ Coulomb 
repulsion is believed to prevent
conventional on-site Cooper pair formation.
This long-standing theoretical dictum seems to
be now supported by experimental evidence, 
obtained in several cuprate materials, for a non-$s$-wave
pairing state, apparently of
$d_{x^2-y^2}$-symmetry \cite{vanharlingen,scalapino}.
Yet, at least one material, the $(Nd,Ce)_2CuO_4$-system,
appears to exhibit a fully gapped 
excitation spectrum, 
suggesting an $s$-wave state within the framework
of more conventional pairing theories
\cite{ncco_exp,syz}.
A $d$-wave pairing state implies
a spatially extended Cooper pair wavefunction
involving electrons paired at 1st neighbor or larger 
lattice distances. 
While impervious to the on-site $U$
repulsion, even non-$s$-wave extended pairing states 
{\it can} be suppressed by the extended 
(1st, 2nd, ... neighbor) part of the electron-electron 
Coulomb repulsion. 
For any proposed microscopic model of the cuprates,
it is therefore of crucial importance to demonstrate that,
firstly, the model can support at least two qualitatively 
different pairing states of, possibly, different symmetry\cite{syz}
and, secondly, that such pairing states are robust against 
on-site {\it and} extended Coulomb repulsions.

In the present paper, we use a combination of diagrammatic
and quantum Monte Carlo (QMC) techniques to show that the charge 
fluctuations in a quasi two-dimensional (2D) extended Hubbard model
give rise to a screening effect which not only
reduces the magnitude, but in fact reverses
the sign of the extended part of the three-dimensional
(3D) screened Coulomb potential $V_S$.
At larger doping, this overscreening becomes so
strong that even the on-site part of $V_S$ 
changes sign and turns attractive.
The overscreening is intrinsically a large-$U$ effect
of the Hubbard system at finite doping 
near band-filling $1\over2$.
Over a doping range of about $10-20\%$, 
$|V_S|$ is much weaker than the bare $U$
and thus offers the prospect of developing {\it controlled}
perturbative expansions in terms of $V_S$, rather than $U$.
Taken as an effective pairing potential, $V_S$ supports
$d_{x^2-y^2}$ pairing, with a pairing strength maximum 
in the $\sim 10-20\%$ doping range, and, in close proximity, 
$s$-wave pairing at larger doping.

We start from an extended Hubbard Hamiltonian
\begin{eqnarray}
H &=& \sum_{j,\ell} 
    \Big(
   {1\over2} V(r_{j\ell}) n_j n_\ell
   -\small\sum_{\sigma} t_{j\ell}
    c_{j\sigma}^\dagger c_{\ell\sigma}
    \Big) 
\nonumber\\
  &\equiv& H_V + H_t
\;,
\;\label{eq:model}
\end{eqnarray}
with $c_{j,\sigma}^\dagger$ creating an 
electron of spin $\sigma=\uparrow,\downarrow$
at $Cu$-site $r_j$ in a 3D crystal 
of stacked $CuO_2$ layers,
$n_j=\sum_\sigma c_{j\sigma}^\dagger c_{j\sigma}$ and 
$r_{j\ell}=r_j-r_\ell$. $H_t$ includes only an in-plane
1st neighbor hybridization $t$ and the chemical
potential $\mu$. The 3D Coulomb potential
\begin{equation}
V(r) = U\delta_{r,0} +{e^2\over \epsilon_B |r|} (1-\delta_{r,0})
\equiv U\delta_{r,0} + V_e(r)
\label{eq:coulomb}
\end{equation}
includes the on-site ($r=0$) repulsion $U$ and an 
extended $1/|r|$-part, $V_e$, with a dielectric constant 
$\epsilon_B$ to account for screening by
the insulating background not explicitly included in $H$,
that is "non-Hubbard" electrons in lower
filled bands and, possibly, phonon degrees of freedom.

The basic idea of our approach is to treat some short-range
portion of $V(r)$, denoted by $V_o(r)$, exactly by QMC methods.
The remaining weaker, but long-range
part of $V$, denoted by $V_\ell(r)$, is then handled perturbatively. 
By including only short-range in-plane terms in $V_o$, 
the QMC simulation can be restricted to a single 2D layer for which
we obtain, by QMC, the density correlation function 
\begin{eqnarray}
\chi_o(q,i\omega) &&= 
{1\over N} \sum_{j,\ell}\int_0^\beta d\tau
e^{i\omega\tau-iq\cdot r_{j\ell}}
\langle \Delta n_j(\tau)_o \Delta n_\ell(0)_o \rangle_o  
\nonumber\\
&&\equiv -P_o(q,i\omega)[1-V_o(q)P_o(q,i\omega)]^{-1}
\label{eq:chi_o}
\end{eqnarray}
at wavevectors $q$, Matsubara frequencies $i\omega$
and temperature $T\equiv 1/\beta$ for lattice size $N$
with $\Delta n_j\equiv n_j-\langle n_j\rangle_o$. 
Here, $\langle...\rangle_o$ and $...(\tau)_o$ denote,
respectively, thermal averaging and imaginary-time 
evolution with respect to
the QMC Hamiltonian $H_o\equiv H_{V_o} + H_t$.
$P_o$ is
the corresponding irreducible polarization 
insertion 
\cite{mahan}. 

The exact screened potential $V_S$,
irreducible polarization 
insertion $P$ and density correlation 
function $\chi$
of the full Hamiltonian (\ref{eq:model}) 
are related by\cite{mahan}
\begin{eqnarray}
V_S(q,i\omega) 
&&= [1-V(q) P(q,i\omega)]^{-1} V(q)
\nonumber\\
&&= V(q) - V(q)\chi(q,i\omega) V(q)
\;,\label{eq:v_s}
\end{eqnarray}
where $V(q)$ denotes the lattice Fourier sum over $V(r)$.
Our essential approximation is to replace the exact 
$P$ of the full Hamiltonian $H$ by $P_o$, 
extracted from the QMC results for $\chi_o$ via Eq.\ref{eq:chi_o}.
All renormalizations of $P$ due to $V_o$ are thus included
exactly, to all orders in $V_o$. Renormalizations due
to the weaker long-range part $V_\ell$ are neglected in $P$,
but approximately included in $V_S$ via Eq.~\ref{eq:v_s}.
Note here that Eqs.~\ref{eq:chi_o},~\ref{eq:v_s} 
for $V_S$, $P$ and $P_o$ are based
on the diagrammatic expansion in the {\it charge} representation
where the $U$-term is written as ${U\over2}\sum_j n_j^2$,
rather than the more familiar {\it spin} representation 
$U\sum_j n_{j\uparrow}n_{j\downarrow}$. While both are equivalent when
summed exactly to all orders, the former, as we will show,
has some crucial advantages for approximate diagram resummations.

As a simple cuprate, we consider $La_{2-x}Sr_xCuO_4$,
with a body-centered tetragonal ($bct$) model crystal structure,
in-plane lattice constant $a=3.80\AA$, inter-layer spacing
$d=6.62\AA$ \cite{jorgensen}, $t=0.35eV$ and $U=8t$ \cite{sf}. 
Using standard determinant QMC methods,
$\chi_o$ is simulated with 
up to $2\times10^7$ MC sweeps and
typically
$\lesssim 0.5\%$ statistical error on $6\times6$, $8\times 8$
and $10\times 10$ 2D lattices
with periodic boundary conditions, taking
$V_o(r)=U\delta_{r,0}$, at
$\beta\equiv1/T$ up to $\beta t=10$, 
with imaginary-time step 
$\Delta\tau\equiv\beta/L\!\leq\!0.0625t^{-1}$
where $L$ is the Trotter number.

To estimate $\epsilon_B$ from the measured 
long-wavelength external field dielectric
function $\epsilon_{ex}(\omega)|_{q\to0}$ in 
the undoped, insulating $La_2CuO_4$ \cite{eps_expa,eps_expb}
we calculate \cite{eps_th}
\begin{equation}
{1\over\epsilon_{ex}(q,i\omega)} 
= {1\over\epsilon_B}
\Big[
1 - {4 \pi e^2\over \epsilon_B {\cal V}_c |q|^2} \chi(q,i\omega)
\Big]
\;,\label{eq:e_ex}
\end{equation}
at the smallest $q$-vectors with non-zero in-plane component
available on our finite model lattices and $i\omega=0$,
with ${\cal V}_c$ denoting the 3D unit cell volume,
$\chi=-P[1-V(q)P]^{-1}$ from (\ref{eq:v_s}) and $P\!\cong\!P_o$.
In the undoped Hubbard system at temperatures 
$T\!\geq\!0.1t$,
we find that $\epsilon_{ex}$ from (\ref{eq:e_ex})
varies roughly linearly with $\epsilon_B$.
The excess 
$\Delta\epsilon\equiv\epsilon_{ex}-\epsilon_B \cong 0.75-0.85$,
that is, the Hubbard electrons' contribution to the 
dielectric screening, is approximately independent of 
$\epsilon_B$ for $\epsilon_B \geq 3$.
{}From the measured dielectric function $\epsilon_{ex}$
of undoped $La_2CuO_4$ in the static limit $\omega\to0$,
$\epsilon_0\equiv\epsilon_{ex}(0)\cong 30$\cite{eps_expa},
we thus estimate 
$\epsilon_B=\epsilon_{ex}-\Delta\epsilon\cong 29$.
However, this includes a large, in fact, 
dominant phonon contribution \cite{emin}.
The purely electronic dielectric screening is observed
at frequencies $\omega_\infty\sim 0.5-1$eV,
well above the phonon spectrum $\Omega_{ph}\lesssim 0.1$eV, 
but still well below the electronic Mott-Hubbard charge gap
$\Delta_{MH}\sim 1.5-2{\rm eV}$, where
$\epsilon_\infty\equiv\epsilon_{ex}(\omega_\infty)\cong 5$ \cite{eps_expb}.
Hence we get
$\epsilon_B\cong 4.2$
for the purely electronic background screening.
>From the estimated values $\epsilon_B\!\cong\!4.2$ (without phonons)
or even $\epsilon_B\cong 29$ (including phonons), one obtains
a quite substantial 1st neighbor repulsion strength
$V_1=e^2/(\epsilon_B a)\cong0.90{\rm eV}\cong 2.6t$ in the former
and $V_1\cong0.13{\rm eV} \cong 0.37t$ in the latter case.
Thus, $V_e(r)$ could severely suppress
spatially extended pairing potentials \cite{t_j_model}.

\begin{figure}
\hskip5mm \epsfig{file=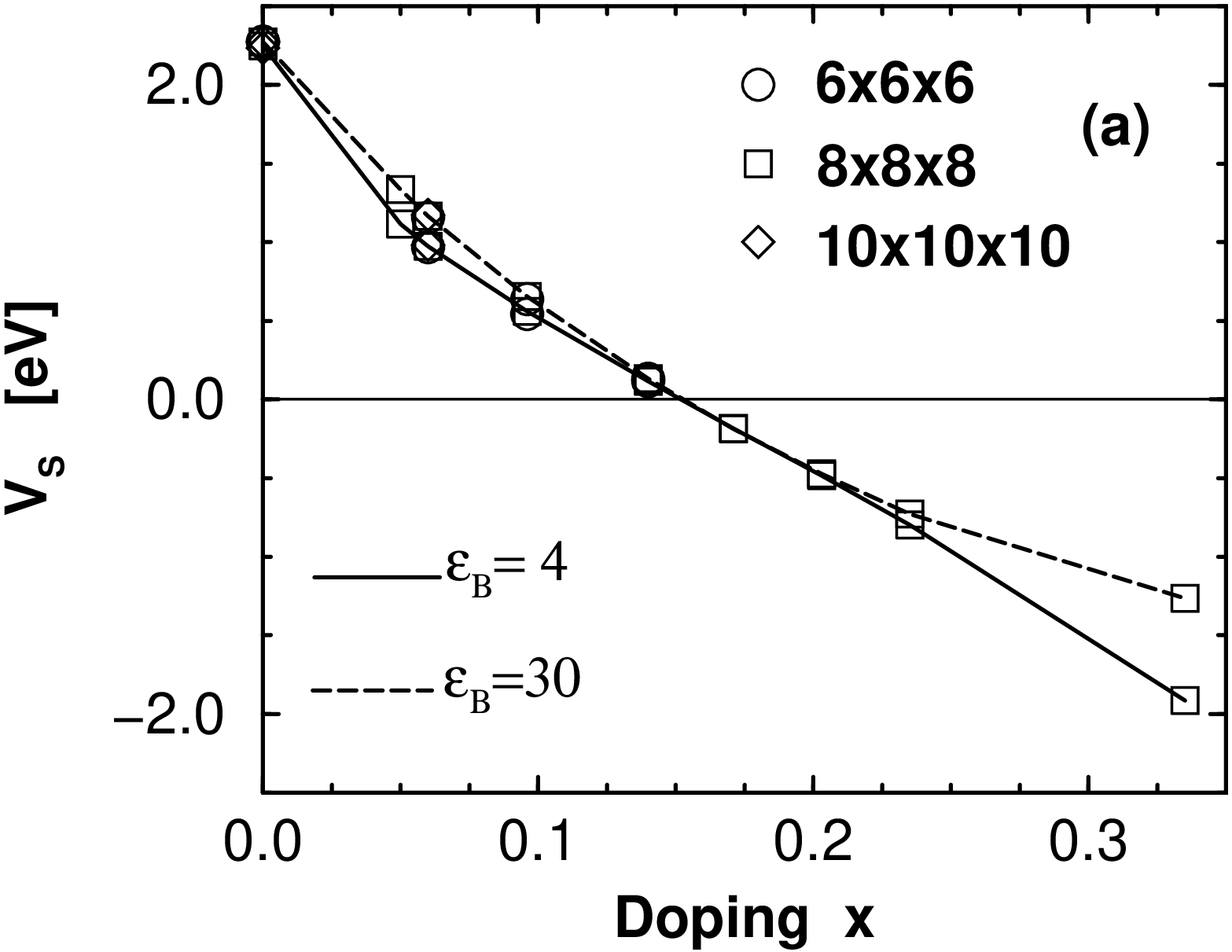,angle=-90,width=7.5cm}  \vskip5mm

\hskip5mm \epsfig{file=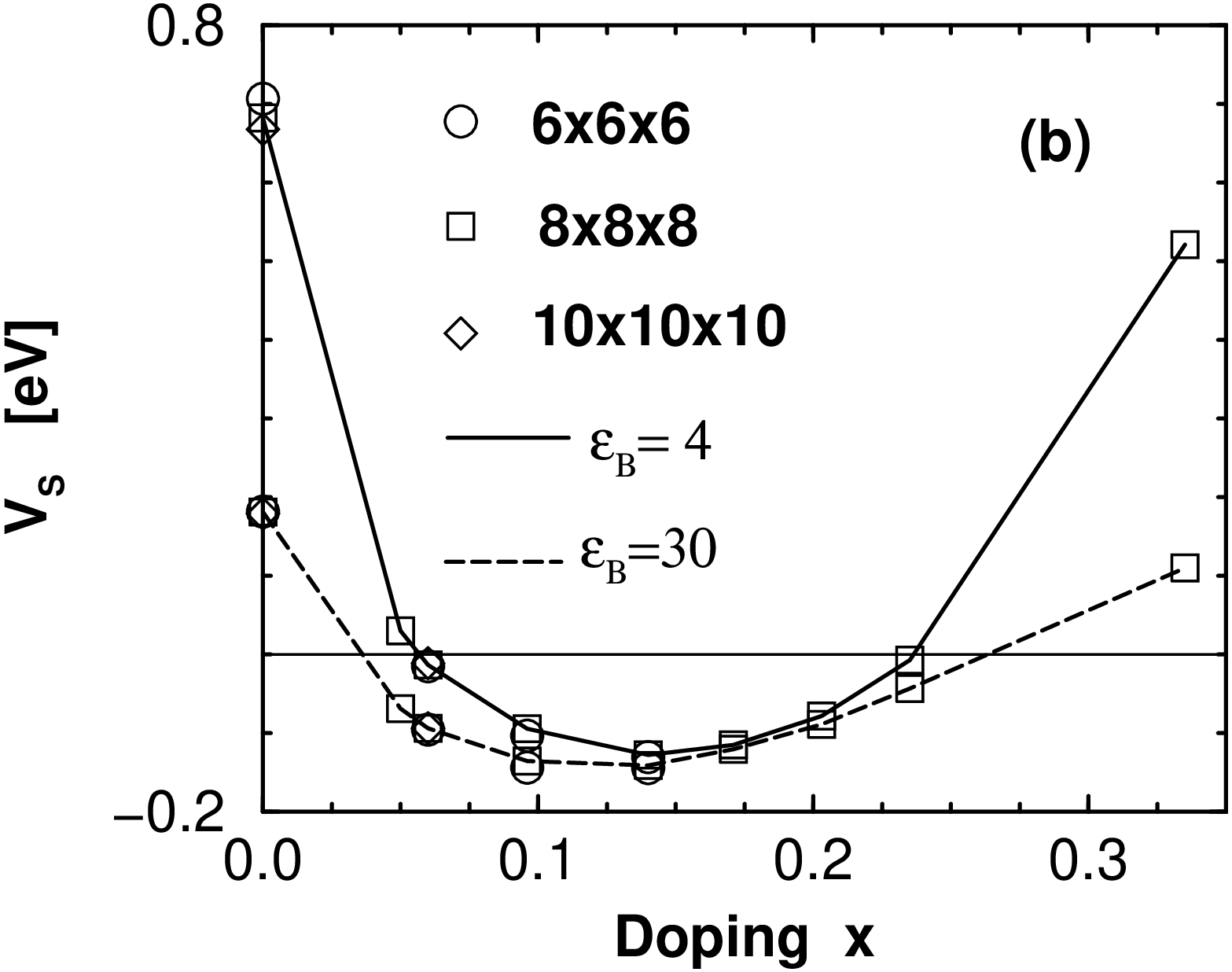,angle=-90,width=7.5cm}  \vskip5mm

\hskip5mm \epsfig{file=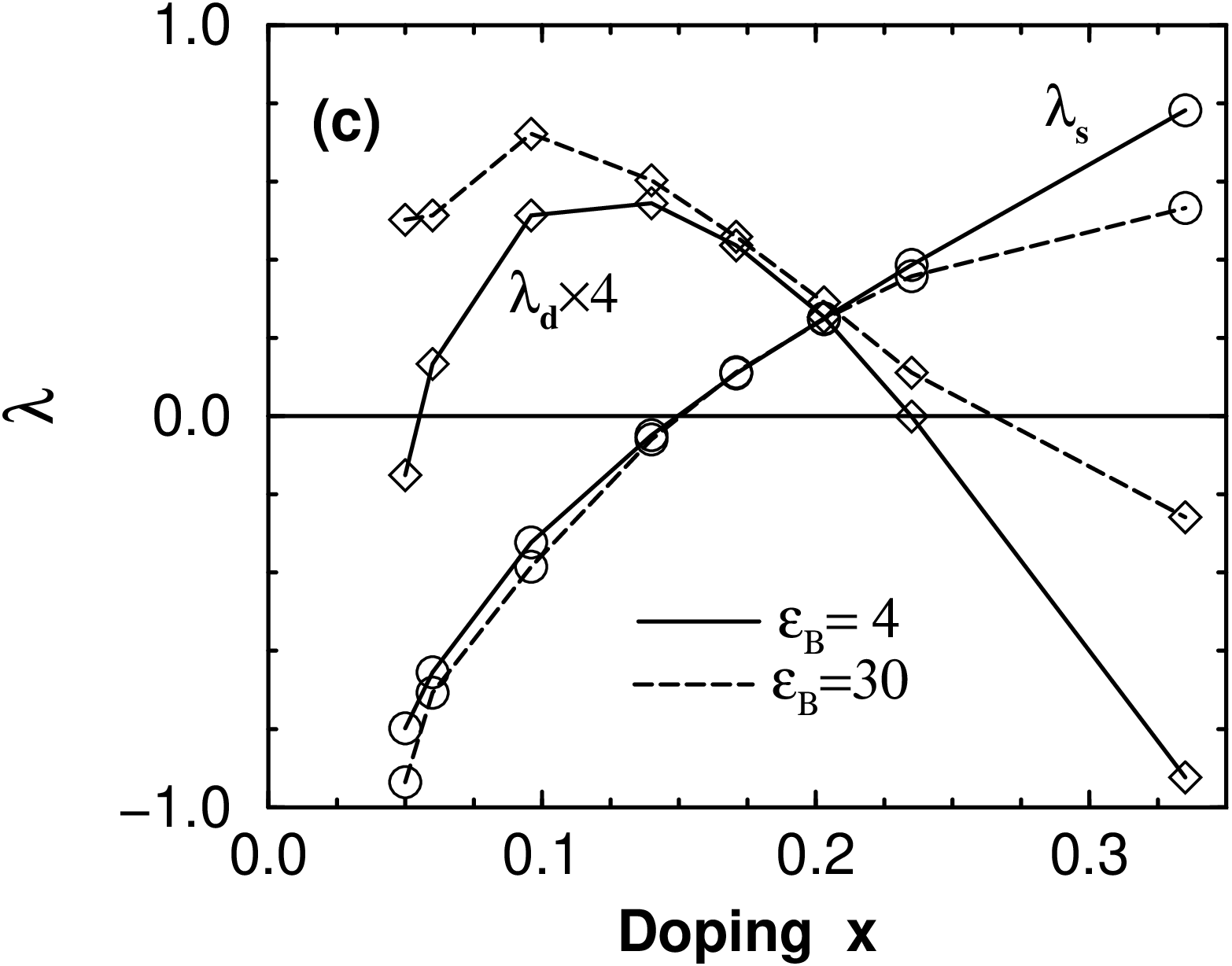,angle=-90,width=7.5cm}  \vskip5mm
\caption{
Screened Coulomb potential 
$V_S(r)\!\equiv\!V_S(r,i\omega=0)$ at
(a) on site ($r=0$) and (b) at in-plane 1st neighbor 
lattice vectors $r$ and (c) Eliashberg $\lambda$-parameters 
for in-plane 1st neighbor $d_{x^2-y^2}$ and 
on-site/1st neighbor $s$-wave pairing, all plotted
vs. hole doping concentration $x=1-\langle n_j \rangle$
at $\beta t\equiv t/T=3.0$ and $\Delta\tau t=0.0375$
for $\epsilon_B=4$ and $30$.
In (a) and (b), results are for $La_2CuO_4$ with $bct$ lattice sizes 
$6\times6\times6$, $8\times8\times8$ and $10\times10\times10$,
with estimated statistical uncertainties in $V_s$ of less than
$0.01{\rm eV}$.
In (c), results are based on $8\times8\times8$ data for $V_S$. 
$\lambda_d$ is scaled $\times\!4$ for 
display.}\label{fig:v_s-x}\end{figure}

In Fig.\ref{fig:v_s-x}(a) and (b), we explore how screening
affects the 3D Coulomb potential $V_S$, calculated from 
(\ref{eq:v_s}) at $i\omega\!=\!0$, Fourier transformed
back to $r$-space and plotted {\it vs.} doping concentration 
$x\equiv 1-\langle n_j\rangle$ for 
the on-site ($r\!=\!0$) and in-plane 1st neighbor $r$-vector.
The surprising result in Fig.~\ref{fig:v_s-x}(b)
is that a small amount of doping, of order $5\%$, will
not only suppress the extended $1/|r|$-repulsion for 
$r\neq0$, but will in fact cause a sign change in the 1st neighbor
and similarly (not shown) in the 2nd and 3rd neighbor screened
potential. Thus $V_S(r)$ becomes attractive at
short-range distances. The attraction strength at $r\neq0$ 
shows a pronounced doping dependence, reaches a maximum at 
$x\!\sim\!10-14\%$ and becomes repulsive again
at $x\!\sim\!23-28\%$. 
Even more surprising, as shown in Fig.~\ref{fig:v_s-x}(a),
is the doping dependence of the on-site ($r\!=\!0$) 
potential which is also rapidly suppressed with
increasing $x$ and becomes strongly attractive at larger doping,
near $x\!\cong\!15\%$. Minus sign problems at finite doping 
unfortunately limit our simulations to $T\!\geq\!0.33t$.
However, at least in that temperature regime, we find $|V_S|$ 
to be increasing with decreasing $T$. This suggests that the
overscreening becomes stronger than shown in Fig.~\ref{fig:v_s-x}
at lower $T$.

The presence of a strong Hubbard-$U$ 
{\it and} finite doping density $x\!>\!0$
are crucial for the overscreening.
If one replaces the $V_o$-renormalized $P_o$ in (\ref{eq:v_s})
by, say, the non-interacting ("RPA") 
polarization bubble $P_{RPA}$, one also
obtains a suppression of $V_S(r)$. However, both the
on-site ($r\!=\!0$) and the short-range
extended part ($r\neq 0$) of $V_S$ 
remain repulsive in RPA \cite{liu_levin}.
Likewise, in the undoped large-$U$ system, $V_S(r)$
is only reduced relative to $V(r)$, 
by a roughly $r$-independent factor,
comparable to the ratio
$\epsilon_{ex}/\epsilon_B$,
for $r\neq0$.
Thus, $V_S$ retains a $1/|r|$-dependence and remains repulsive.
This is expected for the screening of a Coulomb potential
in an insulator and confirms the insulating character
of the ${1\over2}$-filled Hubbard system.

To understand the central role of $U$ and finite 
doping $x$ in the overscreening effect, 
consider the pure 2D Hubbard model where $V(r)=V_o(r)=U\delta_{r,0}$, 
our approximation $P\cong P_o$ becomes exact and, from Eq.~\ref{eq:v_s},
\begin{equation}
V_S(r,i\omega)=U\delta_{r,0}-U^2\chi_o(r,i\omega)
\label{eq:v_s-hub}
\end{equation}
with $\chi_o(r,i\omega)$ denoting the Fourier transform
of $\chi_o(q,i\omega)$.
Clearly, the on-site potential $V_S(r\!=\!0,i\omega)$ 
at finite doping {\it must} become attractive for $U\!\to\!\infty$,
since $\chi_o(r\!=\!0,i\omega)$ is always positive
and approaches a non-zero $U$-independent limit
[of, at least, ${\cal O}(x/t)$, by a simple $U\!=\!\infty$
scaling argument] for $U\gg t$.
By contrast, at $1\over2$-filling, 
all charge fluctuations are suppressed, with 
$\chi_o(r,i\omega=0)\!\sim\!{\cal O}(t^2/U^3)$,
and $V_S(r\!=\!0,i\omega\!=\!0)\!\cong\!U$
is repulsive for $U\!\gg\!t$.
For near (1st, 2nd, ...) neighbor $r$'s,
$\chi_o(r,i\omega\!=\!0)$ is negative at small $U$
and becomes positive only at finite doping and only
when $U$ exceeds some doping dependent 
threshold of order several $t$.
The near neighbor attraction in $V_S$ therefore
also requires large $U$ and finite doping.

The results in Fig.~\ref{fig:v_s-x} 
are almost independent of $V_e$ for $\epsilon_B\!\geq\!4$.
(The $\epsilon_B=30$ data are
within $1-2\%$ of the pure Hubbard ($\epsilon_B\to\infty$) results.)
By using spatially truncated versions of $V_e$,
different layered geometries, and more realistic
dielectric background models \cite{eps_th}, 
we have also verified that neither 3D Coulomb interlayer terms,
nor the long-range $1/|r|$-tail in $V_e$, 
nor our particular choice of crystal structure, 
nor local field effects will substantially affect
the overscreening.
To test our approximation, $P\cong P_o$,
we have also carried out QMC simulations with both
$U$- {\it and} a 1st neighbor $V_e$-term included
in $V_o$. Preliminary results
suggest that setting $P\cong P_o$ in (\ref{eq:v_s})
reproduces the main $V_e$-effect on $\chi$, 
but tends to underestimate the attraction in $V_S$.
Thus, the essential features of $V_S$ 
are very robust against extended Coulomb terms.
This result can be rationalized by expanding 
(\ref{eq:v_s}) to leading order in $V_e$. 
The corrections are smaller
than the $U^2\chi_o$-term in (\ref{eq:v_s-hub})
by a factor of order $V_1\chi_o$ for $U\!\gg\!t$ and 
finite $x\!\gtrsim\!t/U$. 
By simple $U\!\to\!\infty$ scaling 
arguments this is of order 
$V_1x/t$,
{\it i.e.} small if $V_1\!\ll\!t/x$.
At sufficiently large $V_1$, i.e. low $\epsilon_B$,
our basic approach {\it does} break down
due to charge density wave instabilities,
signaled by $1/\chi(q,i\omega\!=\!0)\!\to\!0$.
For the present parameter set and doping range,
this happens only for $\epsilon_B\lesssim 1.8$.

A crucial advantage of our diagrammatic expansion
in the charge representation is the large
reduction of the overall strength 
of $V_S$ in the $10-20\%$ doping range.
This suggests the possibility of 
carrying out {\it controlled}, self-consistent weak-coupling
expansions in which the fully screened $V_S$, 
rather than the bare $V$ or $U$, serves as the small parameter. 
As a first step in that direction,
we have explored possible $V_S$-induced or -enhanced
superconducting pairing instabilities, using
the standard Eliashberg-McMillan (EM) approach \cite{mcmillan}.
A convenient measure of the pairing strength of $V_S$
are the dimensionless EM $\lambda$-parameters,
defined in terms of the Fermi surface (FS) "expectation values" 
of $V_S(k-k',i\omega\!=\!0)$ 
for relevant Cooper pair trial wavefunctions $\eta(k)$
in electron momentum ($k$-) space, as
described in detail in Refs.~\cite{mcmillan}. 

In Fig. 1(c), we show the EM parameters 
$\lambda_s$ for on-site $s$-wave 
(and, identically, for in-plane 
1st neighbor $s$-wave),
and $\lambda_d$, for in-plane 1st neighbor $d_{x^2-y^2}$ pairing, 
with respective pair wavefunctions $\eta_s(k)\equiv 1$ 
and $\eta_d(k)=\cos(ak_x)-\cos(ak_y)$.
For the required FS integrals, we have interpolated our 3D
$V_S(q,i\omega)$ from the finite-lattice $8\!\times\!8\!\times\!8$ 
onto a $200\!\times\!200\!\times\!200$ $q$-mesh.
At low doping, the dominant 
attractive ($\lambda\!>\!0$) channel 
is $d_{x^2-y^2}$ with $\lambda_d$ reaching a maximum of
$\sim\!0.15-0.17$ near $x\!\sim 10-14\%$. 
$\lambda_s$ is repulsive at low doping, 
but becomes strongly attractive
at larger doping $x\gtrsim15\%$. 
Thus, as expected on symmetry grounds,
$\lambda_s$ and $\lambda_d$
reflect the doping dependence 
of the on-site and 1st neighbor 
attraction $V_S$ shown in 
Fig.~\ref{fig:v_s-x}(a) and (b), respectively.
The $\lambda$-values for near-neighbor pair wavefunctions
of other symmetries ($p$, $d_{xy}$, $g$) are
small compared to $\lambda_s$ and $\lambda_d$. 

The spectral weight of $\chi(q,i\omega)$ extends
up to values
$\Omega_\chi\!\sim\!8-10t$ \cite{preuss}.
In the EM analysis \cite{mcmillan}, this "boson" energy scale,
together with $\lambda$, determines the
superconducting $T_c$, roughly as 
$T_c\sim\Omega_\chi\exp(-1/\lambda)$.
Because of the large $\Omega_\chi$-scale, it may be
possible to achieve high $T_c$'s
even at moderate coupling values $\lambda\!<\!1$.

It is quite possible that near-neighbor,
$d$-wave-attractive charge correlations
$\chi(r\!\neq\!0,i\omega)>0$ are very closely related
to short-range antiferromagnetic spin correlations in 
Hubbard systems near $1\over2$-filling.
The charge fluctuation picture developed here 
may thus provide a description of the physics
in near-${1\over2}$-filled Hubbard systems
which is complementary to that of a spin fluctuation
based approach \cite{scalapino}.
The overscreening of the on-site potential, 
and hence the possibility of 
$s$-wave pairing in the Hubbard model,
is one aspect of this problem 
which is obvious in the former, but difficult to
capture in the latter approach.

In summary, we have studied the effect of screening
on the electron-electron interaction potential
in a quasi-2D Hubbard model for $CuO_2$-layers,
coupled by an extended $1/|r|$ 3D Coulomb repulsion. 
While contributing only a minor portion of the 
total insulating dielectric constant at ${1\over 2}$-filling,
the Hubbard electron system, when doped away from ${1\over2}$-filling,
exhibits a strong overscreening effect which causes the extended part 
of the screened potential to change sign and become attractive,
at 1st and further neighbor distances.
This screened potential gives rise to a pairing attraction
in the $d_{x^2-y^2}$ channel which, as a function
of doping, exhibits a maximum near $\sim\!10-15\%$,
reminiscent of the doping dependence of the superconducting $T_c$
in the cuprates. At larger doping, even the on-site part of the
screened potential becomes attractive and gives rise to
an $s$-wave pairing attraction which increases 
strongly with doping, suggesting the possibility 
of a doping-induced transisition
or cross-over from $d$- to $s$-wave pairing.
The overscreening effect is robust against 
3D extended Coulomb repulsions,
independent of the 3D crystal structure,
and represents intrinsically a {\it charge} fluctuation
aspect of the Hubbard electron system
at large $U$ and finite doping density near $1\over2$-filling.
\acknowledgments
One of us (H.-B.S.) would like to acknowledge discussions
with D. Emin, G. Esirgen, K. Levin and S.K. Sinha.
This work was supported by NSF Grant No. DMR-9215123,
by BMBF (05SB8WWA1), by DFN Contract No. TK 598-VA/D03,
and by computing resources from UCNS, University of Georgia,
HLRZ J\"ulich and HLRS Stuttgart.

\end{document}